\documentclass[twocolumn,american,nofootinbib,preprintnumbers]{revtex4}
\usepackage{ae}
\usepackage{aecompl}
\usepackage[T1]{fontenc}
\usepackage[latin1]{inputenc}
\usepackage{amsmath}
\usepackage{graphicx}
\usepackage{amssymb}

\makeatletter
\usepackage{psfrag}

\usepackage{babel}
\makeatother
\begin{document}

\preprint{YITP-08-40}

\title{A volume-weighted measure for eternal inflation}

\author{Sergei Winitzki$^{1,2}$}

\affiliation{$^{1}$Department of Physics, Ludwig-Maximilians University, Munich, Germany}

\affiliation{$^{2}$Yukawa Institute of Theoretical Physics, Kyoto University, Kyoto, Japan}

\begin{abstract}
I propose a new volume-weighted probability measure for cosmological {}``multiverse''
scenarios involving eternal inflation. The {}``reheating-volume (RV) cutoff''
calculates the distribution of observable quantities on a portion of the reheating
hypersurface that is conditioned to be finite. The RV measure is gauge-invariant,
does not suffer from the {}``youngness paradox,'' and is independent of initial
conditions at the beginning of inflation. In slow-roll inflationary models with
a scalar inflaton, the RV-regulated probability distributions can be obtained
by solving nonlinear diffusion equations. I discuss possible applications of
the new measure to {}``landscape'' scenarios with bubble nucleation. As an
illustration, I compute the predictions of the RV measure in a simple toy landscape.
\end{abstract}
\maketitle

\section{Introduction}

In cosmological scenarios such as the {}``recycling universe''~\cite{Garriga:1997ef}
or the string-theoretic landscape~\cite{Bousso:2000xa,Susskind:2003kw,Douglas:2003um},
the fundamental theory does not predict with certainty the values of {}``constants
of nature,'' such as the effective cosmological constant and particle masses.
The cosmological observables may vary significantly between different causally
disconnected regions of the spacetime. Hence one may only hope to obtain the
\emph{probability distribution} of the cosmological observables. Heuristically,
one would like to compute probability distributions of the cosmological parameters
as measured by an observer randomly located in the spacetime. However, eternal
inflation produces an infinite volume in which possible observers may find themselves.
Thus one runs into an immediate difficulty of defining a {}``randomly chosen''
location within a noncompact space.

Observers may appear only after reheating; the physics after reheating is tightly
constrained by current experimental knowledge. The average number of observers
produced in any freshly-reheated spatial domain is a function of cosmological
parameters in that domain. Calculating that function is, in principle, a well-defined
astrophysical problem that does not involve any infinities. Here I focus on
obtaining the probability distribution of cosmological observables at reheating.

\begin{figure}
\begin{centering}\psfrag{t}{$t$}\psfrag{x}{$x$}\includegraphics[width=0.5\columnwidth,keepaspectratio]{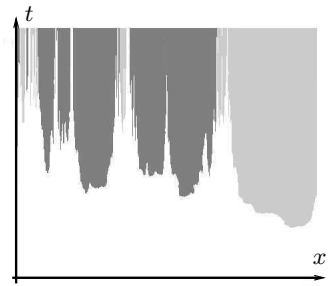}\par\end{centering}

\caption{A 1+1-dimensional slice of the spacetime in an eternally inflating universe
(numerical simulation in Ref.~\cite{Vanchurin:1999iv}). Shades of different
color represent different regions where reheating took place. The reheating
surface is the line separating the white (inflating) domain and the shaded domains.\label{cap:reheating-surface-1} }
\end{figure}

The set of all spacetime points where reheating takes place is a spacelike three-dimensional
hypersurface~\cite{Borde:1993xh,Vilenkin:1995yd,Creminelli:2008es} called
the {}``reheating surface.'' The hallmark feature of eternal inflation is
that a \emph{finite}, initially inflating spatial 3-volume typically gives rise
to a reheating surface having an \emph{infinite} 3-volume, and even (potentially)
to infinitely many causally disconnected pieces of the reheating surface, each
having an infinite 3-volume (see Fig.~\ref{cap:reheating-surface-1}). This
feature of eternal inflation is at the root of several technical and conceptual
difficulties known collectively as the {}``measure problem'' (see Refs.~\cite{Guth:2000ka,Aguirre:2006ak,Winitzki:2006rn,Vilenkin:2006xv,Guth:2007ng,Linde:2007nm}
for reviews and discussions of this problem).

To visualize the measure problem, it is convenient to consider an initial inflating
spacelike region $S$ of horizon size (an {}``$H$-region'') and the portion
$R\equiv R(S)$ of the reheating surface that corresponds to the comoving future
of $S$. If the 3-volume of $R$ were finite, the volume-weighted average of
any observable quantity $Q$ at reheating would be defined simply by averaging
$Q$ over $R$,\begin{equation}
\left\langle Q\right\rangle \equiv\frac{\int_{R}Q\sqrt{\gamma}d^{3}x}{\int_{R}\sqrt{\gamma}d^{3}x},\label{eq:Q average}\end{equation}
where $\gamma$ is the induced metric on the 3-surface $R$. However, in the
presence of eternal inflation%
\footnote{Various equivalent conditions for the presence of eternal inflation were examined
in more detail in Refs.~\cite{Winitzki:2001np,Winitzki:2005ya} and \cite{Creminelli:2008es}.
Here I adopt the condition that $X(\phi)$ is nonzero for all $\phi$ in the
inflating range.%
} the 3-volume of $R$ is infinite with a nonzero probability $X(\phi_{0})$,
where $\phi=\phi_{0}$ is the initial value of the inflaton field at $S$. The
function $X(\phi_{0})$ can be computed in slow-roll inflationary models where
typically $X(\phi_{0})\approx1$~\cite{Winitzki:2001np}. The geometry and
topology of the infinite reheating surface is quite complicated. For instance,
the reheating surface contains infinitely many future-directed {}``spikes''
around never-thermalizing comoving worldlines called {}``eternally inflating
geodesics''~\cite{Winitzki:2001np,Winitzki:2005fy,Vanchurin:2006xp}. In a
spacetime diagram such as Fig.~\ref{cap:reheating-surface-1}, these spikes
reach out to a timelike infinity since the eternally inflating geodesics never
intersect the reheating surface. It is known that the set of spikes has a well-defined
fractal dimension that can be computed in the Fokker-Planck approach~\cite{Winitzki:2001np}. 

For an infinite volume of $R$, the straightforward average~(\ref{eq:Q average})
of a fluctuating quantity $Q(x)$ over a noncompact reheating surface $R$ is
mathematically undefined. The average $\left\langle Q\right\rangle $ can be
computed only after imposing a \emph{volume} \emph{cutoff} on the reheating
surface in some way. A volume cutoff (or a {}``measure'') is, in effect, a
physically motivated prescription that makes volume averages $\left\langle Q\right\rangle $
well-defined. 

Volume cutoffs are usually implemented by restricting the consideration to a
finite portion $\mathcal{V}$ of the reheating domain $R$. After imposing a
cutoff, one computes the {}``regularized'' distribution $p(Q|\mathcal{V})$
of an observable $Q$ by gathering statistics over a large but finite volume
$\mathcal{V}$. The final probability distribution $p(Q)$ is then defined as
\[
p(Q)\equiv\lim_{\mathcal{V}\rightarrow\infty}p(Q|\mathcal{V}),\]
provided that the limit exists. A cutoff prescription is a specific choice of
the compact subset $\mathcal{V}$ and of the way $\mathcal{V}$ approaches infinity
when the cutoff is removed. It has been found early on (e.g.~\cite{Linde:1993xx,Vilenkin:1995yd})
that $p(Q)$ depends sensitively on the choice of the cutoff. Without a natural
mathematical definition of the measure, one judges a cutoff prescription viable
if its predictions are not obviously pathological. Possible pathologies include
the dependence on choice of spacetime coordinates~\cite{Winitzki:1995pg,Linde:1995uf},
the {}``youngness paradox''~\cite{Linde:1994gy,Vilenkin:1998kr,Tegmark:2004qd,Bousso:2007nd},
and the {}``Boltzmann brain'' problem~\cite{Page:2006dt,Linde:2006nw,Vilenkin:2006qg,Page:2006ys,Bousso:2007nd,Gott:2008ii}.

The presently viable cutoff proposals fall into two rough classes that may be
designated as {}``worldline-based'' and {}``volume-based'' measures; a more
fine-grained classification of measure proposals can be found in Refs.~\cite{Aguirre:2006ak,Vanchurin:2006xp}.
Here I propose a new volume-based measure called the {}``reheating-volume
(RV) cutoff.''

\section{Reheating-volume cutoff}

In the RV cutoff, the reheating surface is not being restricted to an artificially
chosen domain. Instead, one simply selects only those initial regions $S$ that,
\emph{by rare chance}, evolve into compact reheating surfaces $R$ having a
finite, fixed volume $\text{Vol}(R)=\mathcal{V}$. The ensemble ${\cal E}_{\mathcal{V}}$
of such initial regions $S$ is a nonempty subset of the ensemble ${\cal E}$
of all initial regions $S$. The volume-weighted probability distribution $p(Q|{\cal E}_{\mathcal{V}})$
of a cosmological observable $Q$ in the ensemble ${\cal E}_{\mathcal{V}}$
can be determined through ordinary sampling of the values of $Q$ over the finite
volume $\mathcal{V}$. The RV cutoff defines the probability distribution $p(Q)$
as the limit of $p(Q|{\cal E}_{\mathcal{V}})$ at $\mathcal{V}\rightarrow\infty$,
provided that the limit exists.

To develop an approach for practical computations in the RV cutoff, let us first
consider the probability density $\rho(\mathcal{V};\phi_{0})d\mathcal{V}$ of
having \emph{finite} volume $\text{Vol}(R)\in[\mathcal{V},\mathcal{V}+d\mathcal{V}]$
of the reheating surface $R$ that results from a single $H$-region with initial
value $\phi=\phi_{0}$. This distribution is normalized to the probability of
the event $\text{Vol}(R)<\infty$, namely\begin{equation}
\int_{0}^{\infty}\rho(\mathcal{V};\phi_{0})d\mathcal{V}=\textrm{Prob}\left(\text{Vol}(R)<\infty\right)=1-X(\phi_{0}).\end{equation}
 The probability density $\rho(\mathcal{V};\phi_{0})$ is nonzero since $X(\phi_{0})<1$.
I call this $\rho(\mathcal{V};\phi_{0})$ the {}``finitely produced reheated
volume'' (FPRV) distribution. This and related distributions constitute the
mathematical basis of the RV cutoff. 

Below I will use the Fokker-Planck (or {}``diffusion'') formalism to derive
equations from which the FPRV distributions can be in principle computed for
models of slow-roll inflation with a single scalar field. Generalizations of
$\rho(\mathcal{V};\phi_{0})$ to multiple-field or non-slow-roll models are
straightforward since the Fokker-Planck formalism is already developed in those
contexts (e.g.~\cite{Helmer:2006tz,Tolley:2008na}).

Let us now define the FPRV distribution for some cosmological observable $Q$
at reheating. Consider the probability density $\rho(\mathcal{V},\mathcal{V}_{Q_{R}};\phi_{0},Q_{0})$,
where $\phi_{0}$ and $Q_{0}$ are values of $\phi$ and $Q$ in the initial
$H$-region, $\mathcal{V}$ is the total reheating volume, and $\mathcal{V}_{Q_{R}}$
is the portion of the reheating volume where the observable $Q$ has a particular
value $Q_{R}$. The distribution $\rho(\mathcal{V},\mathcal{V}_{Q_{R}};\phi_{0},Q_{0})$
as a function of $\mathcal{V}_{Q_{R}}$ at fixed and large $\mathcal{V}$ is
sharply peaked around a mean value $\left\langle \left.\mathcal{V}_{Q_{R}}\right|_{\mathcal{V}}\right\rangle $
corresponding to the average volume of regions with $Q=Q_{R}$ within the total
reheated volume $\mathcal{V}$. Hence, although the full distribution $\rho(\mathcal{V},\mathcal{V}_{Q_{R}};\phi_{0},Q_{0})$
could be in principle determined, it suffices to compute the mean value $\left\langle \left.\mathcal{V}_{Q_{R}}\right|_{\mathcal{V}}\right\rangle $.
One can then expect that the limit \begin{equation}
p(Q_{R})\!\equiv\!\lim_{\mathcal{V}\rightarrow\infty}\!\frac{\left\langle {\!\left.\mathcal{V}_{Q_{R}}\right|}_{\mathcal{V}}\right\rangle }{\mathcal{V}}\!=\!\lim_{\mathcal{V}\rightarrow\infty}\!\!\frac{\!\int_{0}^{\infty}\!\!\rho\left(\mathcal{V},\mathcal{V}_{Q};\phi_{0},Q_{0}\right)\!\mathcal{V}_{Q}d\mathcal{V}_{Q}}{\mathcal{V}\,\mbox{Prob}(\text{Vol}(R)=\mathcal{V})}\label{eq:pQ RV}\end{equation}
exists and is independent of $\phi_{0}$ and $Q_{0}$. (Below I will justify
this statement more formally.) The function $p(Q_{R})$ is then interpreted
as the mean fraction of the reheated volume where $Q=Q_{R}$. In this way, the
RV cutoff yields the volume-weighted distribution for any cosmological observable
$Q$ at reheating. 

\begin{figure}
\begin{centering}\psfrag{x}{$x$}\psfrag{t}{$t$}\includegraphics[width=0.5\columnwidth,keepaspectratio]{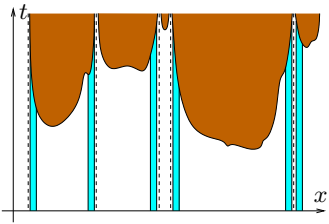}\par\end{centering}

\caption{A schematic representation of the ensemble $E_{\mathcal{V}}$ in comoving coordinates
$\left(t,x\right)$. Lightly shaded vertical strips represent the comoving future
of various initial $H$-regions from $E_{\mathcal{V}}$; dark shades represent
reheated domains; the boundary of the dark-shaded domains is the reheating surface.
Vertical dashed lines are the eternally inflating comoving worldlines that never
cross the reheating surface. The 3-volumes of the reheating surfaces in the
comoving future of the pictured $H$-regions are large but finite because these
$H$-regions are located near eternal worldlines but do not contain any such
worldlines.\label{cap:reheating-surface-2} }
\end{figure}

To obtain a more visual picture of the RV cutoff, consider a large number of
initially identical $H$-regions having different evolution histories to the
future. A small subset of these initial $H$-regions will generate \emph{finite}
reheating surfaces. An even smaller subset of $H$-regions will have the total
reheated volume equal to a given value $\mathcal{V}$. Conditioning on a finite
value $\mathcal{V}$ of the reheating volume, one obtains a well-defined statistical
ensemble $E_{\mathcal{V}}$ of initial $H$-regions. For large $\mathcal{V}$,
the ensemble $E_{\mathcal{V}}$ can be pictured as a set of initial $H$-regions
that happen to be located very close to some eternally inflating worldlines
but do not actually contain any such worldlines (see Fig.~\ref{cap:reheating-surface-2}).
In this way, the ensemble $E_{\mathcal{V}}$ samples the total reheating surface
near the {}``spikes'' where an infinite reheated 3-volume is generated from
a finite initial 3-volume. It is precisely near these {}``spikes'' that one
would like to sample the distribution of observable quantities along the reheating
surface. Therefore, one expects that the ensemble $E_{\mathcal{V}}$ (in the
limit of large $\mathcal{V}$) provides a representative sample of the infinite
reheating surface, despite the small probability of the event $\text{Vol}(R)=\mathcal{V}$.
In this sense, the ensemble $E_{\mathcal{V}}$ at large $\mathcal{V}$ is designed
to yield a controlled approximation to the infinite reheating surfaces $R$. 

The RV cutoff proposed here has several attractive features. By construction,
the RV cutoff is coordinate-invariant; indeed, only the intrinsically defined
3-volume within the reheating surface $R$ is used, rather than the 3-volume
within a coordinate-dependent 3-surface. The results of the RV cutoff are also
independet of initial conditions. This independence is demonstrated more formally
below and can be understood heuristically as follows. The evolution of regions
$S$ conditioned on a large (but finite) value of $\text{Vol}(R)$ is dominated
by trajectories that spend a long time in the high-$H$ inflationary regime
and thereby gain a large volume. These trajectories forget about the conditions
at their initial points and establish a certain equilibrium distribution of
values of $Q$ on the reheating surface. Hence, one can expect that the distribution
of observables within the reheating domain $R$ will be independent of the initial
conditions in $S$.

The {}``youngness paradox'' arises in some volume-based prescriptions because
$H$-regions with delayed reheating are rewarded by an exponentially large additional
volume expansion. However, the RV measure groups together the $H$-regions that
produce \emph{equal} final reheated volume $V$; a delay in reheating is not
rewarded but suppressed by the small probability of a quantum fluctuation at
the end of inflation. Therefore, most of these $H$-regions have {}``normal''
slow-roll evolution before reheating. For this reason, the youngness paradox
is absent in the RV measure. A more explicit calculation confirming this conclusion
will be given in Sec.~\ref{sec:RV-cutoff-for-landscape}.

\section{RV cutoff in slow-roll inflation}

As a first specific application, I implement the RV measure in a slow-roll inflationary
model with a scalar inflaton $\phi$ and the action\begin{equation}
S=\int\left[-\frac{R}{16\pi G}+\frac{1}{2}\left(\partial_{\mu}\phi\right)^{2}-V(\phi)\right]\sqrt{-g}d^{4}x.\end{equation}
If the energy density is dominated by the potential energy $V(\phi)$, the Hubble
expansion rate $H$ is approximately given by \begin{equation}
H\approx\sqrt{\frac{8\pi G}{3}V(\phi)}.\end{equation}
In the stochastic approach to inflation,%
\footnote{See Refs.~\cite{Vilenkin:1983xq,Starobinsky:1986fx,Goncharov:1987ir} for early
works and Refs.~\cite{Linde:1993xx} and \cite{Winitzki:2006rn} for pedagogical
reviews.%
} the semiclassical dynamics of the field $\phi$ averaged over an $H$-region
is a superposition of a deterministic motion with velocity \begin{equation}
\dot{\phi}=v(\phi)\equiv-\frac{1}{4\pi G}H_{,\phi}\end{equation}
 and a random walk with root-mean-squared step size\begin{equation}
\sqrt{\left\langle \delta\phi\right\rangle ^{2}}=\frac{H(\phi)}{2\pi}\equiv\sqrt{\frac{2D(\phi)}{H(\phi)}},\quad D\equiv\frac{H^{3}}{8\pi^{2}},\label{eq:D def}\end{equation}
during time intervals $\delta t=H^{-1}$. A convenient description of the evolution
of the field at time scales $\delta t\lesssim H^{-1}$ is\begin{equation}
\phi(t+\delta t)=\phi(t)+v(\phi)\delta t+\xi(t)\sqrt{2D(\phi)\delta t},\label{eq:phi delta t}\end{equation}
where $\xi(t)$ is (approximately) a {}``white noise'' variable, \begin{equation}
\left\langle \xi\right\rangle =0,\quad\left\langle \xi(t)\xi(t')\right\rangle =\delta(t-t'),\end{equation}
which is statistically independent for different $H$-regions~\cite{Winitzki:1999ve}.
This stochastic process describes the evolution $\phi(t)$ along a single comoving
worldline. For convenience, we assume that inflation ends in a given $H$-region
when $\phi$ reaches a fixed value $\phi=\phi_{*}$. 

Consider an ensemble of initial $H$-regions $S_{1}$, $S_{2}$, ..., where
the inflaton field $\phi$ is homogeneous and has value $\phi=\phi_{0}$ within
the inflationary regime. Following the spacetime evolution of the field $\phi$
in each of the regions $S_{j}$ along comoving geodesics, we arrive at reheating
surfaces $R_{j}$ where $\phi=\phi_{*}$. Most of the surfaces $R_{j}$ will
have infinite 3-volume; however, some (perhaps small) subset of $S_{j}$ will
have finite $R_{j}$. The $\phi_{0}$-dependent probability, denoted $\bar{X}(\phi_{0})\equiv1-X(\phi_{0})$,
of having a finite volume of $R_{j}$ is a solution of the gauge-invariant equation~\cite{Winitzki:2001np}
\begin{equation}
D(\phi)\bar{X}_{,\phi\phi}+v(\phi)\bar{X}_{,\phi}+3H(\phi)\bar{X}\ln\bar{X}=0,\label{eq:X equ}\end{equation}
with the boundary conditions $\bar{X}(\phi_{*})=1$ and $\bar{X}(\phi_{\text{Pl}})=1$
at reheating and at Planck boundaries. While $\bar{X}(\phi)\equiv1$ is always
a solution of Eq.~(\ref{eq:X equ}), the existence of a nontrivial solution
with $0<\bar{X}(\phi)<1$ indicates the possibility of eternal inflation. The
gauge invariance of Eq.~(\ref{eq:X equ}) is manifest since a change of the
time variable, $\tau(t)\equiv\int^{t}T(\phi)dt$, results in dividing the three
coefficients $D,v,H$ by $T(\phi)$~\cite{Winitzki:1995pg}, which leaves Eq.~(\ref{eq:X equ})
unchanged.

The probability distribution $\rho(\mathcal{V};\phi_{0})$ can be found by considering
a suitable generating function. Let us define the generating function $g(z;\phi_{0})$
by \begin{equation}
g(z;\phi_{0})\equiv\left\langle e^{-z\mathcal{V}}\right\rangle _{\mathcal{V}<\infty}\equiv\int_{0}^{\infty}\negthickspace e^{-z\mathcal{V}}\rho(\mathcal{V};\phi_{0})d\mathcal{V}.\label{eq:g def}\end{equation}
(Note that the formal parameter $z$ has the dimension of inverse volume. The
parameter $z$ can be made dimensionless by a trivial rescaling which we omit.)
The function $g(z;\phi_{0})$ is analytic in $z$ and has no singularities for
$\textrm{Re}\, z\geq0$. Moments of the distribution $\rho(\mathcal{V};\phi_{0})$
are determined as usual through derivatives of $g(z;\phi_{0})$ in $z$ at $z=0$,
while $\rho(\mathcal{V},\phi_{0})$ itself can be reconstructed through the
inverse Laplace transform of $g(z;\phi_{0})$ in $z$. 

The generating function $g(z;\phi)$ has the following multiplicative property:
For two statistically independent $H$-regions that have initial values $\phi=\phi_{1}$
and $\phi=\phi_{2}$ respectively, the sum of the (finitely produced) reheating
volumes $\mathcal{V}_{1}+\mathcal{V}_{2}$ is distributed with the generating
function \begin{equation}
\left\langle e^{-z(\mathcal{V}_{1}+\mathcal{V}_{2})}\right\rangle \!=\!\left\langle e^{-z\mathcal{V}_{1}}\right\rangle \!\left\langle e^{-z\mathcal{V}_{2}}\right\rangle \!=\! g(z;\phi_{1})g(z;\phi_{2}).\label{eq:g multiplicative}\end{equation}
This multiplicative property is the only assumption in the derivation of Eq.~(\ref{eq:X equ})
in Ref.~\cite{Winitzki:2001np}. Hence, $g(z;\phi)$ satisfies the same equation
(we drop the subscript 0 in $\phi_{0}$),\begin{equation}
Dg_{,\phi\phi}+vg_{,\phi}+3Hg\ln g=0.\label{eq:g equ}\end{equation}
The boundary condition at $\phi_{*}$ is $g(z;\phi_{*})=e^{-zH^{-3}(\phi_{*})}$
since an $H$-region with $\phi=\phi_{*}$ is already reheating and has volume
$H^{-3}(\phi_{*})$. The boundary condition at the Planck boundary $\phi_{\text{Pl}}$
(or other boundary where the effective field theory breaks down) is {}``absorbing,''
i.e.~we assume that regions with $\phi=\phi_{\text{Pl}}$ disappear and never
generate any reheating volume: $g(z;\phi_{\text{Pl}})=e^{z\cdot0}=1$. Note
that the variable $z$ enters Eq.~(\ref{eq:g equ}) as a parameter and only
through the boundary conditions. At $z=0$ the solution of Eq.~(\ref{eq:g equ})
is $g(0;\phi)=\bar{X}(\phi)$. Explicit approximate solutions of Eq.~(\ref{eq:g equ})
can be obtained using the methods developed in Ref.~\cite{Winitzki:2001np}. 

Let us now consider FPRV distributions of cosmological parameters $Q$. The
generating function for the distribution $\rho(\mathcal{V},\mathcal{V}_{Q_{R}};\phi,Q)$
discussed above is \begin{equation}
\tilde{g}(z,q;\phi,Q)\equiv\!\iint\! e^{-z\mathcal{V}-q\mathcal{V}_{Q}}\rho(\mathcal{V},\mathcal{V}_{Q};\phi,Q)d\mathcal{V}d\mathcal{V}_{Q}.\end{equation}
The equation for $g(z,q;\phi,Q)$ is derived similarly to Eq.~(\ref{eq:g equ})
and is of the form\begin{equation}
D_{\phi}\tilde{g}_{,\phi\phi}+D_{Q}\tilde{g}_{,QQ}+v_{\phi}\tilde{g}_{,\phi}+v_{Q}\tilde{g}_{,Q}+3H\tilde{g}\ln\tilde{g}=0,\label{eq:g equ 2 fields}\end{equation}
where $D_{\phi},D_{Q},v_{\phi},v_{Q}$ are the suitable kinetic coefficients
representing the {}``diffusion'' and the mean {}``drift velocity'' of $\phi$
and $Q$. The boundary condition at $\phi=\phi_{*}$ is \begin{equation}
\tilde{g}(z,q;\phi_{*},Q)=\exp\left[-\left(z+q\delta_{QQ_{R}}\right)H^{-3}(\phi_{*})\right],\end{equation}
 where we use the delta-symbol defined by $\delta_{QQ_{R}}=1$ if $Q$ belongs
to a narrow interval $\left[Q_{R},Q_{R}+dQ\right]$ and $\delta_{QQ_{R}}=0$
otherwise. 

To obtain the distribution~(\ref{eq:pQ RV}), we need to compute the average
$\left\langle \left.\mathcal{V}_{Q_{R}}\right|_{\mathcal{V}}\right\rangle $
at fixed $\mathcal{V}$. We define the auxiliary generating function \begin{equation}
h(z;\phi,Q)\!\equiv\!\left\langle \mathcal{V}_{Q_{R}}e^{-z\mathcal{V}}\right\rangle \!_{\mathcal{V}<\infty}\!=-\tilde{g}_{,q}(z,q=0;\phi,Q).\end{equation}
Note that $\tilde{g}(z,q=0;\phi,Q)=g(z;\phi)$. The equation for $h(z;\phi,Q)$
then follows from Eq.~(\ref{eq:g equ 2 fields}),\begin{equation}
D_{\phi}h_{,\phi\phi}+D_{Q}h_{,QQ}+v_{\phi}h_{,\phi}+v_{Q}h_{,Q}+3H\left(\ln g+1\right)h=0.\label{eq:h equ}\end{equation}
This \emph{linear} equation contains as a coefficient the function $g(z;\phi)$,
which is the solution of Eq.~(\ref{eq:g equ}). The boundary condition for
Eq.~(\ref{eq:h equ}) is \begin{equation}
h(z;\phi_{*},Q)=e^{-zH^{-3}(\phi_{*})}H^{-3}(\phi_{*})\delta(Q-Q_{R}).\end{equation}
 Here we can use the ordinary $\delta$-function instead of the symbol $\delta_{QQ_{R}}$
because the $\delta$-function enters linearly into the boundary condition.
An appropriate rescaling of the distribution $h$ by the factor $dQ$ is implied
when we pass from $\delta_{QQ_{R}}$ to $\delta(Q-Q_{R})$.

Finally, the expectation value $\left\langle \left.\mathcal{V}_{Q_{R}}\right|_{\mathcal{V}}\right\rangle $
at a fixed $\mathcal{V}$ and the limit~(\ref{eq:pQ RV}) can be found using
the inverse Laplace transform of $h(z;\phi,Q)$ in $z$. 

The computation just outlined allows one, in principle, to obtain quantitative
predictions from the RV measure. Further details and a more direct computational
procedure will be given elsewhere~\cite{Winitzki:2008ph}. Presently, let us
analyze the limit $\mathcal{V}\rightarrow\infty$ in qualitative terms. The
function $p(Q_{R};\mathcal{V})$ is expressed as {[}cf.~Eq.~(\ref{eq:pQ RV})]
\begin{equation}
p(Q_{R};\mathcal{V})\!\equiv\!\frac{\left\langle \left.\mathcal{V}_{Q_{R}}\right|_{\mathcal{V}}\right\rangle }{\mathcal{V}}\!=\!\frac{\int_{-\text{i}\infty}^{\text{i}\infty}\! e^{z\mathcal{V}}h(z;\phi,Q)dz}{\mathcal{V}\int_{-\text{i}\infty}^{\text{i}\infty}\! e^{z\mathcal{V}}g(z;\phi)dz}.\label{eq:pQ RV ans}\end{equation}
The asymptotic behavior of the inverse Laplace transform of $h(z;\phi,Q)$ at
large $\mathcal{V}$ is determined by the locations of the singularities of
$h(z;\phi,Q)$ in the complex $z$ plane. The dominant asymptotics of the inverse
Laplace transform are of the form $\propto\exp(z_{*}\mathcal{V})$, where $z_{*}$
is the singularity with the smallest $|\mbox{Re}\, z_{*}|$. It can be shown
that solutions of Eqs.~(\ref{eq:g equ}) and (\ref{eq:h equ}) cannot diverge
at \emph{finite} values of $\phi$ or $Q$. Thus $g(z;\phi)$ and $h(z;\phi,Q)$
cannot have $\phi$- or $Q$-dependent singularities in $z$. Moreover, the
function $g(z;\phi)$ cannot have pole-like singularities in $z$; the only
possible singularities are branch points where the function $g(z;\phi)$ is
finite but a derivative with respect to $z$ diverges. Furthermore, derivatives
$\partial_{z}^{n}h$ satisfy linear equations with coefficients depending on
the derivatives $\partial_{z}^{n-1}g(z;\phi)$, which diverge at the singularities
of $g$. Hence the singularities of $\partial_{z}^{n}h$ in the $z$ plane coincide
with those of $g(z;\phi)$. For these reasons the limit in Eq.~(\ref{eq:pQ RV})
exists and is independent of the initial values $\phi,Q$. A more detailed analysis
justifying these statements will be given in Ref.~\cite{Winitzki:2008ph}.

\section{RV measure for a landscape\label{sec:RV-cutoff-for-landscape}}

The {}``landscape'' scenarios where transitions between metastable vacuum
states occur via bubble nucleation, promise to explain the values of presently
observed cosmological parameters, such as the effective cosmological constant
$\Lambda$. For this reason it is important to be able to apply the RV measure
to landscape-type scenarios and, in particular, to compute the relative abundances
of different bubble types. 

In the terminology of Ref.~\cite{Garriga:2005av}, {}``terminal bubbles''
are those with nonpositive value of $\Lambda$. No further transitions are possible
from such bubbles because bubbles with $\Lambda<0$ rapidly collapse while bubbles
with $\Lambda=0$ do not support tunneling instantons. The RV measure, as presently
formulated, can be used directly for comparing the abundances of \emph{terminal}
bubbles. (Extending the RV prescription to non-terminal bubble types is certainly
possible but is delegated to a future publication.)

It was shown in Ref.~\cite{Bousso:2007nd} that the bubble volume calculations
may use the simplifying {}``square bubble'' approximation, which neglects
the effects of bubble wall geometry. In this approximation, the evolution of
the landscape is well described by the {}``inflation in a box'' model~\cite{Winitzki:2005ya},
defined as follows. All the vacuum states are labeled by $j=1,...,N$ of which
the terminal states are $j=1,...,N_{T}$. During a time step $\delta t$, an
initial $H$-region of type $j$ expands into $n_{j}\equiv e^{3H_{j}\delta t}$
independent daughter $H$-regions of type $j$. Each of the daughter $H$-regions
then has probability $\Gamma_{jk}$ of changing into an $H$-region of type
$k$ (for convenience we define $\Gamma_{jj}\equiv1-\sum_{k\neq j}\Gamma_{jk}$).
The process is repeated \emph{ad infinitum} for each resulting $H$-region,
except for $H$-regions of terminal types. A newly created $H$-region of terminal
type will admit no further transitions and will not expand (or, perhaps, will
expand only by a fixed amount of slow-roll inflation occurring immediately after
nucleation). This imitates the behavior of anti-de Sitter or Minkowski vacua
that do not admit further bubble nucleations.

To implement the RV cutoff in this model, let us consider the probability $p_{j}(n,n';k)$
of producing a finite total number $n$ of terminal $H$-regions of which $n'$
are of type $j$, starting from one initial $H$-region of (nonterminal) type
$k$. A generating function for this distribution can be defined by \begin{equation}
g_{j}(z,q;k)\equiv\negmedspace\sum_{n,n'=0}^{\infty}\negmedspace z^{n}q^{n'}p_{j}(n,n';k)\equiv\left\langle z^{n}q^{n'}\right\rangle _{n<\infty}.\end{equation}
One can show that this generating function satisfies the following system of
nonlinear algebraic equations,\begin{equation}
g_{j}^{1/n_{k}}(z,q;k)=\sum_{i=1}^{N_{T}}\Gamma_{ki}zq^{\delta_{ij}}+\negmedspace\sum_{i=N_{T}+1}^{N}\negmedspace\Gamma_{ki}g_{j}(z,q;i),\label{eq:g equ landscape}\end{equation}
I will merely sketch the derivation of Eq.~(\ref{eq:g equ landscape}), which
is similar to the equations for generating functions used in the theory of branching
processes (see e.g.~the book~\cite{AthreyaNey:1972:BP} for a mathematically
rigorous presentation). The generating function $g_{j}$ satisfies a multiplicative
property analogous to Eq.~(\ref{eq:g multiplicative}). This property applies
to the $n_{k}$ independent daughter $H$-regions created by expansion from
an $H$-region of type $k$. Therefore, $g_{j}(z,q;k)$, which is the expectation
value of $z^{n}q^{n'}$ in an initial $H$-region of type $k$, is equal to
the product of $n_{k}$ expectation values of $z^{n}q^{n'}$ in the $n_{k}$
daughter $H$-regions (which may be of different types). The latter expectation
value is given by the right-hand side of Eq.~(\ref{eq:g equ landscape}). This
yields Eq.~(\ref{eq:g equ landscape}) after raising both sides to the power
$1/n_{k}$.

If the generating functions $g_{j}$ are known, the distribution $p_{j}(n,n';k)$
can be recovered by computing derivatives of $g_{j}(z,q;k)$ at $z=0$ and $q=0$.
Further, the mean fraction of $H$-regions of type $j$ at fixed total number
$n$ of terminal $H$-regions is found as \begin{equation}
p(j|n)\equiv\frac{\left\langle \left.n'\right|_{n}\right\rangle }{n}=\frac{\partial_{z}^{n}\partial_{q}g_{j}(z=0,q=1;k)}{n\,\partial_{z}^{n}g_{j}(z=0,q=1;k)}.\label{eq:p n' n lim}\end{equation}
Then the RV cutoff \emph{defines} the probability of terminal type $j$, among
all the possible \emph{terminal} types, through the limit \begin{equation}
p(j)\equiv\lim_{n\rightarrow\infty}p(j|n)=\lim_{n\rightarrow\infty}\frac{\partial_{z}^{n}\partial_{q}g_{j}(z=0,q=1;k)}{n\,\partial_{z}^{n}g_{j}(z=0,q=1;k)},\label{eq:p n'n lim 1}\end{equation}
similarly to Eq.~(\ref{eq:pQ RV}). Again one expects that the limit exists
and is independent of the initial bubble type, as long as the initial bubble
is not of terminal type.

As a specific example requiring fewer calculations, let us consider a toy model
with only three bubble types. There is one de Sitter ($\Lambda>0$) vacuum labeled
$j=3$ that can decay into two possible anti-de Sitter terminal bubbles labeled
$j=1$ and $j=2$. The growth rate $n_{3}$ and the nucleation probabilities
$\Gamma_{31}$ and $\Gamma_{32}$ are assumed known. To mimick interesting features
of the landscape, let us also assume that there is a period of slow-roll inflation
inside the bubbles 1 and 2, generating respectively $N_{1}$ and $N_{2}$ additional
$e$-folds of inflationary expansion after nucleation. Hence, the model is determined
by the parameters $n_{3}$, $\Gamma_{31}$, $\Gamma_{32}$, $N_{1}$, and $N_{2}$.
For convenience we define $\Gamma_{33}\equiv1-\Gamma_{31}-\Gamma_{32}$.

We now perform the calculations for the RV cutoff in this simple model. There
are only two generating functions, $g_{1}(z,q;k)$ and $g_{2}(z,q;k)$, that
need to be considered. The only meaningful initial value is $k=3$ (i.e., the
initial bubble is of de Sitter type) since the two other bubble types do not
lead to eternal inflation. Hence, we will suppress the argument $k$ in $g_{j}(z,q;k)$.
We also need to modify Eq.~(\ref{eq:g equ landscape}) to take into account
the additional expansion inside the terminal bubbles. Let us denote the volume
expansion factors by \begin{equation}
Z_{1}\equiv e^{3N_{1}},\quad Z_{2}\equiv e^{3N_{2}}.\end{equation}
 Then the functions $g_{1}$ and $g_{2}$ are solutions of \begin{align}
g_{1}^{1/n_{3}} & =\Gamma_{31}z^{Z_{1}}q^{Z_{1}}+\Gamma_{32}z^{Z_{2}}+\Gamma_{33}g_{1},\\
g_{2}^{1/n_{3}} & =\Gamma_{31}z^{Z_{1}}+\Gamma_{32}z^{Z_{2}}q^{Z_{2}}+\Gamma_{33}g_{2}.\end{align}
An explicit solution of these equations is impossible for a general $n_{3}$
(barring the special cases $n_{3}=2,3,4$). Nevertheless, sufficient information
about the limit~(\ref{eq:p n'n lim 1}) can be obtained by the following method.
Introduce the auxiliary function $F(x)$ as the solution $F>0$ of the algebraic
equation\begin{equation}
F^{1/n_{3}}=x+\Gamma_{33}F,\label{eq:F def}\end{equation}
choosing the branch connected to the value $F(0)=0$. (It is straightforward
to see that Eq.~(\ref{eq:F def}), has at most two positive solutions, and
that $F(x)$ is always the smaller solution of the two.) The generating functions
$g_{1}$ and $g_{2}$ are then expressed through $F(x)$ as\begin{align}
g_{1}(z,q) & =F(\Gamma_{31}z^{Z_{1}}q^{Z_{1}}+\Gamma_{32}z^{Z_{2}}),\label{eq:g1 through F}\\
g_{2}(z,q) & =F(\Gamma_{31}z^{Z_{1}}+\Gamma_{32}z^{Z_{2}}q^{Z_{2}}).\label{eq:g2 through F}\end{align}
The limit~(\ref{eq:p n'n lim 1}) involves derivatives of these functions of
very high order with respect to $z$. We note that the function $F(x)$ is analytic;
thus the functions $g_{1}$ and $g_{2}$ are also analytic in $z$. 

To evaluate the high-order derivatives, we need an elementary result from complex
analysis. The asymptotic growth of high-order derivatives of an analytic function
$f(z)$ is determined by the location of the singularities of $f(z)$ in the
complex $z$ plane. For instance, we may expect an expansion around the singularity
$z_{*}$ nearest to $z=0$, such as \begin{equation}
f(z)=c_{0}+c_{1}\left(z-z_{*}\right)^{s}+...,\label{eq:f sing}\end{equation}
where $s\neq0,1,2,...$ is the power of the leading-order singularity, and the
omitted terms are either higher powers of $z-z_{*}$ or singularities at points
$z_{*}^{\prime}$ located further away from $z=0$. The singularity structure~(\ref{eq:f sing})
yields the large-$n$ asymptotics with the leading term\begin{equation}
\left.\frac{d^{n}f}{dz^{n}}\right|_{z=0}\negmedspace\approx c_{1}(-z_{*})^{s}\frac{\Gamma(n-s)}{\Gamma(-s)}z_{*}^{-n}.\label{eq:large-n derivative}\end{equation}
This formula enables one to evaluate large-$n$ limits such as Eq.~(\ref{eq:p n'n lim 1}).

To proceed, we need to determine the location of the singularities of $F(x)$.
Since $F(x)$ is obtained as an intersection of a curve $F^{1/n_{3}}$ and a
straight line $x+\Gamma_{33}F$, there will be a value $x=x_{*}$ where the
straight line is tangent to the curve. At this value of $x$ the function $F(x)$
has a singularity of the type\begin{equation}
F(x)=F(x_{*})+F_{1}\sqrt{x-x_{*}}+O(x-x_{*}),\label{eq:F sqrt}\end{equation}
where $F_{1}$ is a constant that can be easily determined; we omit further
details that will not be required below. The value of $x_{*}$ is found from
the condition that $dF/dx$ diverge at $x=x_{*}$. The value of $dF/dx$ at
$x\neq x_{*}$ is found as the derivative of the inverse function, or by taking
the derivative of Eq.~(\ref{eq:F def}),\begin{equation}
\frac{dF}{dx}=\frac{1}{\frac{1}{n_{3}}F^{\frac{1}{n_{3}}-1}-\Gamma_{33}}.\end{equation}
This expression diverges at the values \begin{align}
F(x_{*}) & =\left(n_{3}\Gamma_{33}\right)^{-\frac{n_{3}}{n_{3}-1}},\\
x_{*} & =\left(n_{3}-1\right)\Gamma_{33}F(x_{*})=\Gamma_{33}\frac{n_{3}-1}{n_{3}^{n_{3}/(n_{3}-1)}}.\end{align}
Note that $\Gamma_{33}\approx1$ and $n_{3}\gg1$, hence $x_{*}$ is a constant
of order 1. 

Rather than compute the limit~(\ref{eq:p n' n lim}) directly, we will perform
an easier computation of the \emph{ratio} of the mean number of bubbles of types
1 and 2 at fixed total number $n$ of terminal bubbles, \begin{equation}
\frac{\left\langle \left.n_{(1)}^{\prime}\right|_{n}\right\rangle }{\left\langle \left.n_{(2)}^{\prime}\right|_{n}\right\rangle }=\frac{\partial_{z}^{n}\partial_{q}g_{1}(z=0,q=1)}{\partial_{z}^{n}\partial_{q}g_{2}(z=0,q=1)}.\label{eq:n ratio lim}\end{equation}
The derivatives $\partial_{q}g_{1}$ and $\partial_{q}g_{2}$ can be evaluated
directly through Eqs.~(\ref{eq:g1 through F})--(\ref{eq:g2 through F}). For
instance, we compute $\partial_{q}g_{1}$ as \begin{equation}
\left.\frac{\partial g_{1}(z,q)}{\partial q}\right|_{q=1}=F^{\prime}(\Gamma_{31}z^{Z_{1}}+\Gamma_{32}z^{Z_{2}})\Gamma_{31}Z_{1}z^{Z_{1}}.\label{eq:dg1 dq}\end{equation}
It is clear that the functions $\partial_{q}g_{1}$ and $\partial_{q}g_{2}$
have a singularity at $z=z_{*}$ corresponding to the singularity $x=x_{*}$
of the function $F(x)$, where $z_{*}$ is found from the condition\begin{equation}
\Gamma_{31}z_{*}^{Z_{1}}+\Gamma_{32}z_{*}^{Z_{2}}=x_{*}.\label{eq:z star equ}\end{equation}
Let us analyze this equation in order to esimate $z_{*}$. If the nucleation
rates $\Gamma_{31}$ and $\Gamma_{32}$ differ by many orders of magnitude,
we may expect that one of the terms in Eq.~(\ref{eq:z star equ}), say $\Gamma_{31}z_{*}^{Z_{1}}$,
dominates. Then the value $z_{*}$ is well approximated by \begin{equation}
z_{*}\approx\left(\frac{x_{*}}{\Gamma_{31}}\right)^{1/Z_{1}}.\label{eq:z star ans}\end{equation}
This approximation is justified if the first term in Eq.~(\ref{eq:z star equ})
indeed dominates, which is the case if\begin{equation}
\left(\frac{\Gamma_{32}}{x_{*}}\right)^{1/Z_{2}}\ll\left(\frac{\Gamma_{31}}{x_{*}}\right)^{1/Z_{1}}.\label{eq:gamma32 cond}\end{equation}
If the reversed inequality holds, we can relabel the bubble types 1 and 2 and
still use Eq.~(\ref{eq:z star ans}). If neither Eq.~(\ref{eq:gamma32 cond})
nor the reversed inequality hold, the approximation~(\ref{eq:z star ans})
for $z_{*}$ can be used only as an order-of-magnitude estimate. To be specific,
let us assume that the condition~(\ref{eq:gamma32 cond}) holds.

We can now compute the ratio~(\ref{eq:n ratio lim}) asymptotically for large
$n$ using Eqs.~(\ref{eq:F sqrt}) and (\ref{eq:large-n derivative}). The
singularities of the functions $g_{1}$ and $g_{2}$ are directly due to the
singularity of the function $F$. Then the dominant singularity structure of
the function~(\ref{eq:dg1 dq}) is found as\begin{equation}
\left.\frac{\partial g_{1}(z,q)}{\partial q}\right|_{q=1}\approx\left.\frac{\Gamma_{31}Z_{1}z^{Z_{1}}}{2\sqrt{\frac{\partial}{\partial z}\left(\Gamma_{31}z^{Z_{1}}+\Gamma_{32}z^{Z_{2}}\right)}}\right|_{z=z_{*}}\frac{F_{1}}{\sqrt{z-z_{*}}}.\end{equation}
This fits Eq.~(\ref{eq:f sing}), where $f(z)\equiv\partial_{q}g_{1}(z,q=1)$
and $s=-1/2$. After canceling the common $n$-dependent factors, we obtain\begin{equation}
\frac{p(1)}{p(2)}=\lim_{n\rightarrow\infty}\frac{\left\langle \left.n_{(1)}^{\prime}\right|_{n}\right\rangle }{\left\langle \left.n_{(2)}^{\prime}\right|_{n}\right\rangle }=\left.\frac{\Gamma_{31}Z_{1}z^{Z_{1}}}{\Gamma_{32}Z_{2}z^{Z_{2}}}\right|_{z=z_{*}}.\label{eq:prob ratio ans}\end{equation}
This is the final probability ratio found by applying the RV cutoff to a toy
model landscape containing two terminal vacua. Substituting the approximation~(\ref{eq:z star ans}),
which assumes the condition~(\ref{eq:gamma32 cond}), we can simplify Eq.~(\ref{eq:prob ratio ans})
to a more suggestive form\begin{equation}
\frac{p(1)}{p(2)}\approx\frac{\Gamma_{31}Z_{1}}{\Gamma_{32}Z_{2}}\left(\frac{\Gamma_{31}}{x_{*}}\right)^{-1+Z_{2}/Z_{1}}.\label{eq:new answer}\end{equation}
We can interpret this as the ratio of nucleation probabilities $\Gamma_{31}/\Gamma_{32}$
times the ratio of volume expansion factors, $Z_{1}/Z_{2}$, times a certain
{}``correction'' factor. As we have seen, the correction factor is actually
a complicated function of all the parameters of the landscape. The correction
factor takes the simple form \begin{equation}
\left(\frac{\Gamma_{31}}{x_{*}}\right)^{-1+Z_{2}/Z_{1}}\label{eq:correction factor}\end{equation}
 only if the condition~(\ref{eq:gamma32 cond}) holds.

We note that the result~(\ref{eq:new answer}) is similar to but does not exactly
coincide with the results obtained in previously studied volume-based measures.
For comparison, the volume-based measures proposed in Refs.~\cite{Garriga:2005av}
and \cite{Linde:2007nm} both yield\begin{equation}
\frac{p(1)}{p(2)}=\frac{\Gamma_{31}Z_{1}}{\Gamma_{32}Z_{2}},\label{eq:old answer}\end{equation}
which is readily interpreted as the ratio of nucleation probabilities enhanced
by the ratio of volume factors. The {}``holographic'' measure~\cite{Bousso:2006ev},
which is not a volume-based measure, gives the ratio \begin{equation}
\frac{p(1)}{p(2)}=\frac{\Gamma_{31}}{\Gamma_{32}}\end{equation}
that does not depend on the number of $e$-folds after nucleation. While the
discrepancy between volume-based and worldline-based measures is to be expected,
the correction factor that distinguishes Eq.~(\ref{eq:new answer}) from Eq.~(\ref{eq:old answer})
is model-dependent and may be either negligible or significant depending on
the particular model. 

As a specific example, consider bubbles that nucleate with equal probability,
$\Gamma_{31}=\Gamma_{32}\ll1$, but have very different expansion factors, $Z_{1}\gg Z_{2}$.
Then the condition~(\ref{eq:gamma32 cond}) holds and the {}``correction''
factor is given by Eq.~(\ref{eq:correction factor}), which is an exponentially
large quantity of order $\Gamma_{31}^{-1}$. Qualitatively this means that the
RV measure rewards bubbles with a larger slow-roll expansion factor even more
than previous volume-based measures.

On the other hand, if $Z_{1}=Z_{2}$ but $\Gamma_{31}\gg\Gamma_{32}$, the {}``correction''
factor disappears and we recover the result found in the other measures.

To conclude, we note that the result~(\ref{eq:prob ratio ans}) does not depend
on the durations of \emph{time} spent during slow-roll inflation inside the
terminal bubbles, but only on the number of $e$-folds gained. This confirms
that the RV measure does not suffer from the youngness paradox. 

So far we were able to apply of the RV measure to the comparison of the abundances
of terminal vacua. One expects that, with some more effort, the RV measure can
be extended to arbitrary observables in landscape models. Further work will
clarify the advantages and limitations of the new measure.

\section*{Acknowledgments}

The author is grateful to Cedric Deffayet, Jaume Garriga, Takahiro Tanaka, and
Alex Vilenkin for valuable discussions. Part of this work was completed on a
visit to the Yukawa Institute of Theoretical Physics (University of Kyoto).
The stay of the author at the YITP was supported by the Yukawa International
Program for Quark-Hadron Sciences.

\bibliographystyle{myphysrev}
\bibliography{EI2}

\end{document}